\documentclass[final]{cvpr}

\usepackage{times}
\usepackage{epsfig}
\usepackage{graphicx}
\usepackage{amsmath}
\usepackage{amssymb}

\DeclareMathOperator*{\argmin}{arg\,min}

\usepackage{booktabs} 
\usepackage{etoolbox}
\usepackage[detect-weight=true,detect-family=true]{siunitx}
\robustify\bfseries
\usepackage{pifont}  
\newcommand{\cmark}{\ding{51}}%
\newcommand{\xmark}{\ding{55}}%
\usepackage{subcaption}

\usepackage{listings}
\usepackage[dvipsnames]{xcolor}
\usepackage[frozencache]{minted}
\newcommand{\cl}[1]{}
\newcommand{\MZ}[1]{}

\usepackage[pagebackref=true,breaklinks=true,colorlinks,bookmarks=false]{hyperref}

\def\cvprPaperID{***} 
\def\confYear{****}

\begin{document}


\title{Pulsar: Efficient Sphere-based Neural Rendering}

\author{
Christoph Lassner$^1$~~~
Michael Zollhöfer$^1$~~~
\\
$^1$Facebook Reality Labs
}

\newlength{\titleup}
\setlength{\titleup}{-1.1cm}

\twocolumn[{%
\renewcommand\twocolumn[1][]{#1}%
\maketitle
\vspace*{\titleup}
\begin{center}
    \centering
    \includegraphics[width=.87\textwidth]{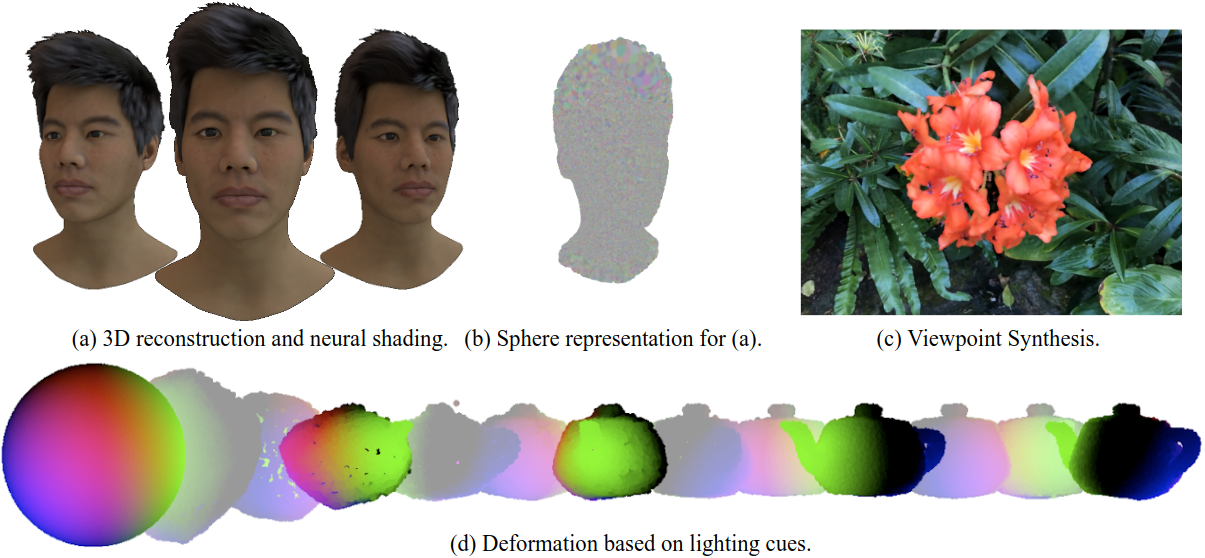}
    \captionof{figure}{
     Pulsar is an efficient sphere-based differentiable renderer that is orders of magnitude faster than competing techniques, modular, and easy-to-use.
     It can be employed to solve a large variety of applications, since it is tightly integrated with PyTorch.
     Using a sphere-based representation, it is possible to not only optimize for color and opacity, but also for positions and radii (a, b, c).
     Due to the modular design, lighting cues can also be easily integrated (d).
     }
\end{center}%
}]

\begin{abstract}
    \vspace*{-0.2cm}
    We propose Pulsar, an efficient sphere-based differentiable renderer that is orders of magnitude faster than competing techniques, modular, and easy-to-use due to its tight integration with PyTorch.
    Differentiable rendering is the foundation for modern neural rendering approaches, since it enables end-to-end training of 3D scene representations from image observations.
    However, gradient-based optimization of neural mesh, voxel, or function representations suffers from multiple challenges, i.e., topological inconsistencies, high memory footprints, or slow rendering speeds.
    To alleviate these problems, Pulsar employs: 1) a sphere-based scene representation, 2) an efficient differentiable rendering engine, and 3) neural shading.
    Pulsar executes orders of magnitude faster than existing techniques and allows real-time rendering and optimization of representations with millions of spheres.
    Using spheres for the scene representation, unprecedented speed is obtained while avoiding topology problems.
    Pulsar is fully differentiable and thus enables a plethora of applications, ranging from 3D reconstruction to general neural rendering.
    \vspace*{0.7cm}
\end{abstract}

\vspace*{\titleup}
\section{Introduction}

Differentiable rendering is the foundation for all modern neural rendering approaches that learn 3D scene representations based on image observations.
Recently, differentiable rendering has empowered a large variety of applications, such as novel-view synthesis \cite{mildenhall2020nerf}, facial reenactment \cite{Thies_DNR}, and 3D reconstruction \cite{liu2019dist}.
Modern neural rendering can be broken up into three components:
1) a 3D neural scene representation, 2) a projection from 3D data to a consistent 2D representation (the \emph{projection step}) and 3) processing the projected data using a statistical model, usually a neural network, to produce the final image (the \emph{neural shading step}).
This strategy combines the strengths of classical rendering and neural networks.
Through the projection step, a consistent geometric representation of the scene is generated, while the neural shading step can produce realistic images through the use of the latest generative neural networks that can approximate complex natural image formation phenomena without having to explicitly model them.

Ideally, such a neural rendering approach can be jointly trained in an end-to-end fashion: a 3D representation of the scene is learned and sent through the projection and shading step.
The resulting image can be compared to ground truth observations to inform an optimization process, not only to improve the generative model in the shading step, but also to jointly learn the representation of the scene and potentially unknown parameters of the projection step.
This process requires the efficient computation of gradients through the complete pipeline in a scalable manner, such that high performance can be obtained even for the geometry of complex and detailed scenes rendered at real-world resolutions.

In this paper, we present \emph{Pulsar}, an efficent, sphere-based, differentiable renderer that is orders of magnitude faster than competing techniques, modular, and easy-to-use due to its tight integration with PyTorch.
Pulsar aims to fulfill all mentioned requirements through a variety of measures, from the design of the scene representation down to low-level data-parallel optimizations, which lead to unprecedented speed for the forward and backward pass.
First, we choose an entirely sphere-based representation of 3D data.
Each sphere is parameterized by its position in space and its radius.
In addition, each sphere has an assigned opacity and can have an arbitrary vector as payload, such as a color or a general latent feature vector.
Image formation is based on a volumetric compositing schema that aggregates the payload in back-to-front order to form the final image or a screen space feature map.
This makes it easy to handle point cloud data from 3D sensors directly, allows for the optimization of the scene representation without problems of changing topology (as they would exist for meshes) and is more efficient for rendering than recent approaches based on volumetric grids or fully-connected networks, since our representation, sparse by design, culls empty space.
In addition, our sphere-based representation eliminates the need for acceleration structures, such as a $k$-d tree or octree, thus naturally can support dynamic scenes.
Additionally, it leads to a well-defined, simple render and blending function that can be differentiated without approximation.
We deliberately keep the illumination computations separate from the geometry projection step as it can be easily handled in a separate step.
Lastly, we integrate Pulsar with the PyTorch~\cite{NEURIPS2019_9015} optimization framework to make use of auto-differentiation and ease the integration with deep learning models.

The strategy described above allows Pulsar to render and differentiate through the image formation for 3D scenes with millions of spheres on consumer graphics hardware.
Up to one million spheres can be rendered and optimized at real-time speed for an image resolution of $1024\times1024$ pixels (the time spent executing C++ code is less than \SI{22}{ms} for rendering and less than \SI{6}{ms} for gradient calculation).
Pulsar supports a generalized pinhole and orthogonal camera model and computes gradients for camera parameters as well as to update the scene representation.
We demonstrate that a large variety of applications can be successfully handled using Pulsar, such as 3D reconstruction, neural rendering, and viewpoint synthesis.
Pulsar is open source and thus will enable researchers in the future to solve a large variety of research problems on their own.
In summary, our main technical contributions are:
\begin{itemize}
    \setlength\itemsep{0em}
    \item A fast, general purpose, sphere-based, differentiable renderer that is tightly integrated in PyTorch and enables end-to-end training of deep models with geometry and projection components.
    \item Pulsar executes orders of magnitude faster than existing techniques and allows real-time rendering and optimization of representations with millions of spheres.
    \item We demonstrate that a large variety of applications can be handled with Pulsar, such as 3D reconstruction, modeling realistic human heads, and novel view synthesis for scenes.
\end{itemize}

\begin{table*}
\begin{center}
  \resizebox{\textwidth}{!}{
  \begin{tabular}{l c c c c c c c}
      \hline
      method & objective & position update & depth update & normal update & occlusion & silhouette change & topology change \\
      \hline
      OpenDR & mesh & \cmark & \xmark & via position change & \xmark & \cmark & \xmark \\
      NMR & mesh & \cmark & \xmark & via position change & \xmark & \cmark & \xmark \\
      Paparazzi & mesh & limited & limited & via position change & \xmark & \xmark & \xmark \\
      Soft Rasterizer & mesh & \cmark & \cmark & via position change & \cmark & \cmark & \xmark \\
      Pix2Vex & mesh & \cmark & \cmark & via position change & \cmark & \cmark & \xmark \\
      Tensorflow Graphics & mesh & \cmark & \cmark & via position change & \cmark & \cmark & \xmark \\
      PyTorch3D & mesh / points & \cmark & \cmark & via position change & \cmark & \cmark & \cmark \\
      DSS & points & \cmark & \cmark & \cmark & \cmark & \cmark & \cmark \\
      \textbf{Pulsar (ours)} & spheres & \cmark & \cmark & via extra channels & \cmark & \cmark & \cmark \\
      \hline
  \end{tabular}
  }
\end{center}
\vspace*{-0.6cm}
\caption{
\label{table_methods}
Feature comparison of generic differentiable renderers (compare to~\cite{yifan2019differentiable}, Tab. 1).
DSS and PyTorch3D are the only other renderers that do not require a mesh-based geometry representation, facilitating topology changes.
In contrast to DSS, Pulsar uses 3D spheres but without normals.
Extra channels can be used to capture and optimize normal information.
}
\vspace*{-0.2cm}
\end{table*}
\vspace{-0.3cm}
\section{Related Work}
We focus our discussion of related work on differentiable rendering and commonly used scene representations, such as textured meshes, voxel grids, (implicit) functions, and point-based representations.
For a comprehensive review of neural rendering, we refer to the recent state of the art report on `Neural Rendering' of Tewari et~al.~\cite{tewari2020}.

\paragraph{Differentiable Rendering}
Differentiable rendering can be understood as a subfield of \emph{inverse graphics}, which has been a part of computer vision research since its early days~\cite{baumgart1974geometric}.
For a summary of the features of current approaches, see Tab.~\ref{table_methods}.
One of the seminal works on differentiable rendering of meshes, including lighting and textures, is OpenDR \cite{Loper:ECCV:2014}.
It is built on top of OpenGL and uses local Taylor expansions and filter operations to find gradients, excluding depth.
OpenDR leverages existing OpenGL infrastructure, but introduces approximations and has a large overhead in running filtering and approximation operations.
Neural Mesh Renderer (NMR)~\cite{kato2018neural} renders meshes using a custom function to address object boundaries.
Paparazzi et al.~\cite{liu2017material} is another mesh renderer that is implemented using image filtering.
Pix2Vex \cite{petersen2019pix2vex} is a mesh renderer that uses soft blending of triangles.
In the same spirit, Liu et al.~\cite{liu2019soft} introduce a probabilistic map of mesh triangles.
They use a soft $z$-buffer to obtain a differentiable representation.
Their rendering function inspired our formulation.
Tensorflow Graphics~\cite{TensorflowGraphicsIO2019} is a differentiable rendering package for Tensorflow~\cite{tensorflow2015-whitepaper} with support for mesh geometry.
Similarly, PyTorch3D~\cite{ravi2020pytorch3d} is a differentiable rendering package for PyTorch and initially focused on mesh rendering.
A recent extension makes point-based rendering available and has been used for creating SynSin~\cite{wiles2019synsin}.
Pulsar executes orders of magnitude faster than these techniques.

\paragraph{Physics-based Differentiable Rendering}
Several renderers aim to be close to the underlying physical processes.
Li et~al.~\cite{azinovic2019inverse,Li:2018:DMC} implement differentiable ray tracers to be able to compute gradients for physics-based rendering effects.
These implementations explicitly model the image formation process in much greater detail, but are significantly slower in execution.
Similarly, the Mitsuba~2 renderer~\cite{NimierDavidVicini2019Mitsuba2} and Difftaichi~\cite{hu2019difftaichi} are physics-based differentiable renderers with slower execution times, but a lot more physical details. Whereas it would be possible to implement Pulsar using Difftaichi or Enoki (Mitsuba's autodiff framework), it would not be possible to implement many of Pulsars optimization strategies.
In this work, we do not focus on physics-based approaches, since our aim is fast real-time differentiable rendering to empower modern neural rendering approaches that approximate natural image formation phenomena without having to explicitly model them.

\paragraph{Scene Representations}
There is a large variety of possible scene presentations from dense voxel grids \cite{choy20163dr2n2,nvs_steve,sitzmann2019deepvoxels,tulsiani2017multi,wu2016learningvox}, multi-plane images \cite{mildenhall2019llff,zhou2018mpi}, meshes \cite{kato2018neural,liu2019soft,CodecAvatars,Loper:ECCV:2014,thies2019neuraltex}, function-based representations \cite{liu2020nsvf, mildenhall2020nerf,schwarz2020graf,sitzmann2019}, to point-based representations (discussed in the next paragraph).
In contrast to using explicit differentiable graphics engines, neural rendering can also be implemented solely through deep learning models.
This is, for example, attempted in~\cite{kulkarni2015deep}.
Implicit functions, such as signed distance fields (SDFs), are a popular representation for geometry.
Recently, fully connected networks \cite{park2019deepsdf} have been used to learn SDFs.
Liu et al.~\cite{liu2019dist} optimize a signed distance function via differentiable sphere tracing.
Similarly, Saito et al.~\cite{saito2019pifu} model humans through an implicit function.
Zeng et~al.~\cite{zeng2020arch} optimize a similar function using a differentiable renderer.
Jiang et~al.~\cite{jiang2019sdfdiff} is implementing differentiable rendering directly for SDFs, including lighting.
RenderNet~\cite{nguyen2018rendernet} is a CNN architecture with a projection unit.
Tulsiani et al.~\cite{lsiTulsiani18} use a layered depth image representation and develop a differentiable renderer for optimizing this representation.
Instead of prescribing a fixed input size or discretizing the underlying 3D scene, Sitzmann et al.~\cite{sitzmann2019} and Mildenhall et al.~\cite{mildenhall2020nerf} represent the scene using the network structure and employ variants of ray-casting to reconstruct images from arbitrary viewpoints.

\paragraph{Point-based Representations}
Insafutdinov et al.~\cite{insafutdinov18pointclouds} propose to work with differentiable point clouds.
They train a CNN to predict the shape and pose of an object in 3D and use an orthographic projection and ray termination probabilities to obtain a differentiable representation.
In contrast to our approach, their method is strongly limited in terms of resolution (they use $128\times 128$ pixel image resolution and only up to \num{16}k points in their paper); this is too low to solve real world tasks.
Yifan et~al.~\cite{yifan2019differentiable} propose a point-based representation with a position and normal parameterization.
Each point is rendered as a small `surface' with position, radius, and normal.
In the examples shown in their paper, they use representations with up to \num{100}k points and report orders of magnitude slower runtime than our approach (\SI{258}{ms} forward and \SI{680}{ms} backward for an image of resolution $256\times 256$ pixels).
Lin et~al.~\cite{lin2018learning} define a renderer for point cloud generation, but only provide gradients for depth values.
Roveri et~al.~\cite{roveri2018network} define a point-based differentiable projection module for neural networks that produces `soft' depth images.
Aliev et~al.~\cite{aliev2019neural}~propose to model room-scale point clouds with a deferred neural rendering step.
SynSin~\cite{wiles2019synsin} and the PyTorch3D point renderer follow a similar approach to ours, but are orders of magnitude slower.
In addition, we employ a different blending function and enable the optimization of the sphere radius.
Furthermore, they use only the first few points per pixel to determine the pixel colors.
We have found this to leads to high frequency artifacts in complex scenes and thus allow for an unlimited number of spheres to contribute to the pixel color (or set a bound based on the minimum contribution) and use only the first few spheres for gradient propagation.

\section{Method}
\begin{figure}
    \centering
    \includegraphics[width=0.48\textwidth]{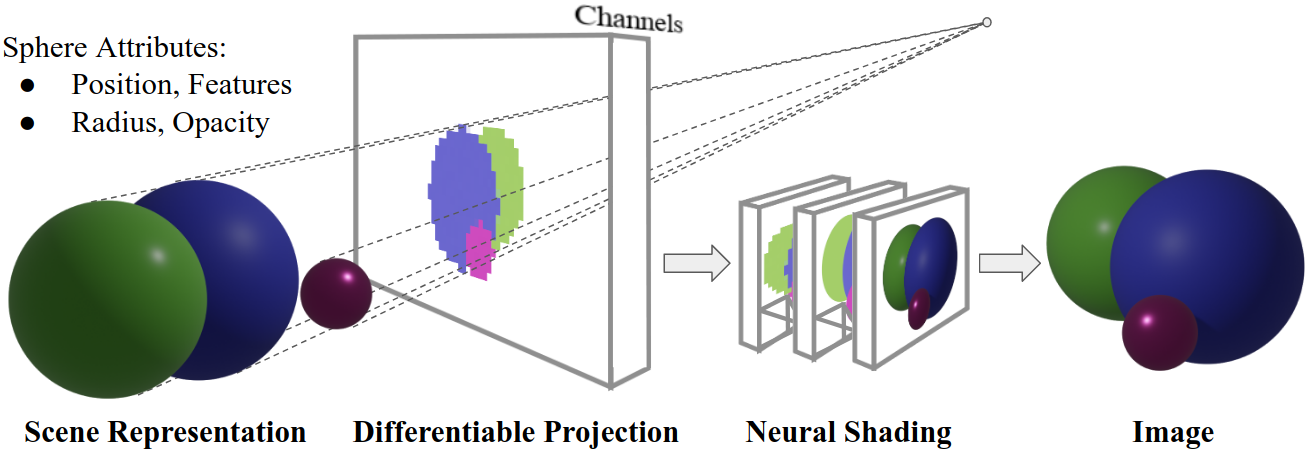}
    \caption{Visualization of the neural rendering pipeline. Pulsar enables a particularly fast differentiable projection step that scales to complex scene representations. The scene representation itself can be produced by a neural network. The channel information can be `latent' and translated to RGB colors in a `neural shading' step.}
    \label{fig:pipeline}
    \vspace*{-0.3cm}
\end{figure}
We are interested in neural rendering approaches that learn a 3D scene representation from a set of $N$ training images $\mathcal{T} = \{(\mathbf{I}_i, \mathbf{R}_i, \mathbf{t}_i, \mathbf{K}_i)\}_{i=1}^N$.
Here, the $\mathbf{I}_i\in \mathbb{R}^{H \times W \times 3}$ are image observations, the $\mathbf{R}_i \in \mathbb{R}^{3\times3}$ are camera rotations, the $\mathbf{t}_i \in \mathbb{R}^{3}$ are camera translations, and the $\mathbf{K}_i\in \mathbb{R}^{3\times3}$ are intrinsic camera parameters~\cite{Hartley:2003}.
The neural rendering task can be split into three stages (see Fig.~\ref{fig:pipeline}): 1)~the scene representation (a traditional mesh model or potentially itself the result of a neural network), 2)~a differentiable rendering operation, and 3)~a neural shading module. Pulsar provides a particularly efficient differentiable rendering operation for scene representations that work with spheres as primitives.
In the following, we provide more details.

\subsection{Sphere-based Scene Representation}
We represent the scene as a set $\mathcal{S}= \{(\mathbf{p}_i, \mathbf{f}_i, r_i, o_i)\}_{i=1}^M$ of $M$ spheres with learned position $\mathbf{p}_i \in \mathbb{R}^3$, neural feature vector $\mathbf{f}_i\in \mathbb{R}^d$, radius $r_i\in \mathbb{R}$, and opacity $o_i\in \mathbb{R}$.
All of these scene properties can be optimized through the differentiable rendering operation.
The neural feature vector $\mathbf{f}_i \in \mathbb{R}^d$ encodes local scene properties.
Depending on its use, it can represent surface color, radiance, or be an abstract feature representation for use in a neural network.
If radiance is directly learned, our scene representation can be understood as an efficient and sparse way to store a neural radiance field \cite{mildenhall2020nerf} by only storing the non-empty parts of space.
The explicit sphere-based scene representation enables us to make use of multi-view and perspective geometry by modeling the image formation process explicitly.

\subsection{Efficient Differentiable Rendering}
Our differentiable renderer implements a mapping $\mathbf{F} = \mathcal{R}(\mathcal{S}, \mathbf{R}, \mathbf{t}, \mathbf{K})$ that maps from the 3D sphere-based scene representation $\mathcal{S}$ to a rendered feature image~$\mathbf{F}$ based on the image formation model defined by the camera rotation $\mathbf{R}$, translation $\mathbf{t}$, and intrinsic parameters $\mathbf{K}$. $\mathcal{R}$ is differentiable with respect to $\mathbf{R}, \mathbf{t}$ and most parts of $\mathbf{K}$, i.e, focal length and sensor size.

\paragraph{Feature Aggregation}
The rendering operation $\mathcal{R}$ has to compute the channel values for each pixel of the feature image $\mathbf{F}$ in a differentiable manner.
To this end, we propose a blending function that combines the channel information based on the position, radius, and opacity of the spheres that are intersected by the camera ray associated with each pixel.
For a given ray, we associate a blending weight $w_i$ with each sphere $i$:
\begin{align}
    w_i = \frac{o_i \cdot d_i \cdot \textcolor{OliveGreen}{\exp}(o_i\cdot\frac{\textcolor{OliveGreen}{z_i}}{\gamma})}{\exp(\frac{\epsilon}{\gamma})+\textcolor{OliveGreen}{\sum_k} o_k\cdot d_k\cdot\textcolor{OliveGreen}{\exp}(o_k\cdot\frac{\textcolor{OliveGreen}{z_k}}{\gamma})}.
    \label{eq:blending}
\end{align}
Similar to Liu et al.~\cite{liu2019soft}, we choose a \textcolor{OliveGreen}{weighed softmax function} of the sphere intersection depth $z_i$ as the basis for our definition.
We employ normalized device coordinates $z_i
\in [0, 1]$ where 0 denotes maximum depth.
A scaling factor $\gamma$ is used to push the representation to be more rigorous with respect to depth.
Small values, such as $\gamma=1^{-5}$, lead to `hard' blending, while large values, such as $\gamma=1$, lead to `soft' blending.
Depending on the quantities that are optimized it makes sense to use different values for gamma.
$\gamma=1$ and $\gamma=1^{-5}$ are the limits we allow to maintain numerical stability.
The additional offset $\exp(\frac{\epsilon}{\gamma})$ is the `weight' for the background color of the scene, for a fixed small constant $\epsilon$.
$d_i$ is the normalized orthogonal distance of the ray to the sphere center.
This distance, since always orthogonal to the ray direction, automatically provides gradients for the two directions that are orthogonal to the ray direction.
We define $d_i = \min(1, \frac{||\vec{d}_i||_2}{R_i})$, where $\vec{d}_i$ is the vector pointing orthogonal from the ray to the sphere center.
Like this, $d_i$ becomes a linear scaling factor in $[0, 1]$.

It is non trivial to integrate opacity into Eq.~\ref{eq:blending} in a differentiable way, since it has to be `soft'.
Assuming there is a per sphere opacity value $o_i$, it could be integrated as a factor into the exponential function, or as another linear scaling factor.
Similar to \cite{zeng2020arch}, we observe that integrating it only as a depth scaling factor often leaves spheres visible in front of the background.
Using it only as a `distance' scaling factor makes depth `override' opacity in most cases and does not lead to appropriate results.
Using it in both places is a feasible and numerically stable solution.

\paragraph{Data-parallel Implementation}
Our renderer is implemented in C++ and CUDA C as a PyTorch extension to make use of the processing power of modern GPUs and naturally integrate the rendering step with machine learning models.
For the data-parallel renderer to be efficient, it is important to find 1)~the right parallelization scheme, 2)~prune unnecessary spheres early on, and 3)~perform early stopping along the rays.
In the following, we discuss the motivation and high-level concepts behind these steps. For a detailed discussion, we refer to the supplemental material.

The most fundamental choice is whether to parallelize the rendering process over pixels or spheres.
Parallelizing over the spheres can be beneficial because of sharing of information for the evaluation of Eq.~\ref{eq:blending}.
However, this approach requires synchronization between threads when writing to the same pixel, which obliterates performance.
The alternative is parallelizing over the pixels.
In this case, it is critical to find a good way to exploit spatial closeness between pixels during the search for relevant spheres.

This allows to reduce the amount of candidate spheres (spheres that could influence the color of a pixel) as quickly as possible.
We propose a tile-based acceleration structure and cooperative filtering across all hardware levels for filtering. An additional benefit of parallelizing over pixels is that the rendering speed is largely independent of sphere sizes.

The sphere intersection search may be stopped early once it has been determined that no sphere with a greater distance can still have a notable impact on the blending function.
For this, we initially sort all spheres device-wide by earliest possible intersection from the camera.
During intersection search, we use Eq.~\ref{eq:blending} and the aggregated weights to find an upper bound for the sphere depth to have at least a certain amount of impact on the blended features.
As soon as the first sphere is tested with a greater depth, the search can be stopped.
Early stopping is of critical importance for the high speed of the renderer and results in better asymptotic performance than that of other renderers.

\subsection{Neural Shading}
The task of the neural shading network $\mathcal{N}$ is to convert the 2D screen space feature map $\mathbf{F}$ to the output color image using a pixel generator network that learns the feature-to-color mapping.
In our tests, we employ either a convolutional U-Net or a per-pixel one-by-one convolutional network.
The higher the capacity of the shading network, the more the approach can overfit.
It is important to find the right trade-off based on the desired application and the available training data.

\subsection{End-to-end Optimization}
We find the best sphere-based neural 3D scene representation $\mathcal{S}^*$ and neural shading model $\mathcal{N}(\bullet;\Theta_{s}^*)$ through gradient-based optimization using Pulsar.
We solve the following end-to-end optimization problem:
$$
\mathcal{S}^*, \Theta_{s}^* = \argmin_{\mathcal{S}, \Theta_{s}}{\sum_{i=0}^{N}{ \big|\big|\mathbf{I}_i - \mathcal{N}(\mathcal{R}(\mathcal{S}, \mathbf{R}_i, \mathbf{t}_i, \mathbf{K}_i); \Theta_s)}\big|\big|_1}.
$$
Note that this optimization finds all parameters based on the 2D image observations without any additional 3D supervision.
Pulsar provides the efficient differentiable renderer $\mathcal{R}$, while all other stages are implemented in a modular manner using PyTorch.
We use ADAM in all experiments to solve this optimization problem.
Whereas this is one straightforward application of the proposed rendering component, it can be used in a variety of settings and integrated in deep learning pipelines in a straightforward and modular way.

\sisetup{detect-all,group-minimum-digits=5}
\begin{table*}
    \begin{center}
        \resizebox{\textwidth}{!}{
        \begin{tabular}{l r r r r}
            \hline
            method & number of points & number of faces & avg. forward time in ms & avg. backward time in ms \\
            \hline
            Soft Rasterizer & \num{15099} & \num{29924} & \num{285} & \num{294}\\
            DSS & \num{15099} & n.a. & \num{215} & \num{179} \\
            PyTorch3D (mesh) & \num{15099} & \num{29924} & \num{104} & \num{80} \\
            PyTorch3D (points) / SynSin & \num{15099} & n.a. & \num{34} & \num{2} \\
            pulsar & \num{15099} & n.a. & \num{14} & \num{1} \\
            pulsar (CUDA only) & \num{15099} & n.a. & \num{3} & \num{1} \\
            \hline
            Soft Rasterizer & \num{233872} & \num{467848} & \num{5032} & \num{5356} \\
            DSS & \num{233872} & n.a. & \num{3266} & \num{3690} \\
            PyTorch3D (mesh) & \num{233872} & \num{467848} & \num{222} & \num{105} \\
            PyTorch3D (points) / SynSin & \num{233872} & n.a. & \num{112} & \num{3} \\
            \textbf{Pulsar (Ours)} & \num{233872} & n.a. & \textbf{\num{21}} & \textbf{\num{2}} \\
            \textbf{Pulsar (CUDA only)} & \num{233872} & n.a. & \textbf{\num{9}} & \textbf{\num{1}} \\
            \hline
        \end{tabular}
        }
    \end{center}
    \vspace*{-0.6cm}
    \caption{
    Runtime performance comparison of state-of-the-art differentiable renderers with PyTorch integration.
    For Pulsar, we measure the performance using the full Python interface (as for the other renderers) as well as the runtime of the CUDA kernel.
    \textit{PyTorch3D (points)} uses a fixed point size for all points and the runtime does not scale well for larger point sizes.
    Pulsar's runtime is largely sphere size agnostic and scales favorably with resolution and number of spheres.
    For example, for 1 million spheres we still measure execution times of less than 33ms (19ms in CUDA) forward and 11ms (4.7ms in CUDA) backward.
    All times are measured on an NVIDIA RTX 2080 GPU at $1000\times 1000$ image resolution.}
    \label{tab:runtimes}
    \vspace*{-0.5cm}
\end{table*}
\sisetup{group-minimum-digits=5}

\section{Results}
In the following, we compare to other differentiable renderers in terms of training and test time performance and demonstrate the power and simplicity of the modular implementation of Pulsar as a PyTorch module.
\vspace*{-0.5cm}
\paragraph{Runtime Performance}

\begin{figure}
    \includegraphics[width=0.46\textwidth]{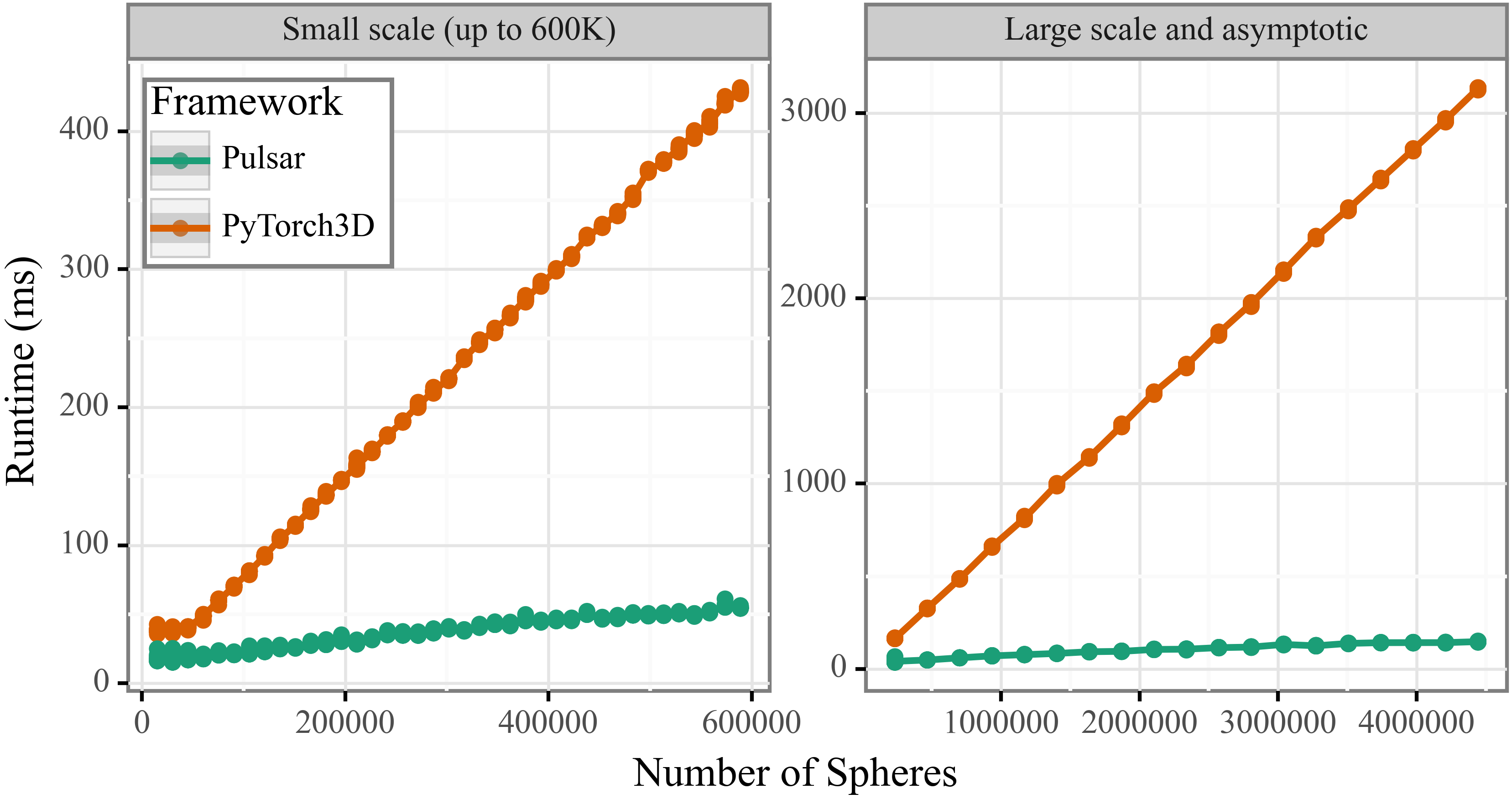}
    \vspace*{-0.3cm}
    \caption{Scaling behavior for PyTorch3D and Pulsar for different numbers of spheres. Whereas PyTorch3D scales almost linearly in terms of number of spheres, Pulsar employs early-stopping and other optimization techniques to reach much better scaling behavior. Benchmarks performed on an NVIDIA RTX 2070 GPU at $1024\times 1024$ resolution.}\label{fig:scaling}
    \vspace*{-0.5cm}
\end{figure}

Rendering speed is even more important for differentiable and neural rendering than for traditional rendering, because it effectively limits the resolution of the images and scene representations that can be processed: the scene is not processed only `once' in a forward pass, but continuously within an optimization loop. Optimizations with millions of spheres, as presented later in this paper, are prohibitively slow to perform with other renderers.
To illustrate this, we compare Pulsar to a large variety of state-of-the-art differentiable rendering approaches in terms of runtime on two scenes of varying complexity.

Pulsar outperforms all current state-of-the-art approaches, including PyTorch3D, by a large margin (see Tab.~\ref{tab:runtimes}), in some comparisons more than two orders of magnitude and at least factor five.
We continued performing measurements for the closest contenders, \textit{PyTorch3D (points)} and Pulsar, to analyze the asymptotic behavior. The other frameworks generally were too slow to create meaningful comparisons on this scale. You can find the results in Fig.~\ref{fig:scaling}. For PyTorch3D, we used \texttt{points\_per\_pixel=5} to achieve a close match of conditions. The PyTorch3D sphere renderer already reaches a runtime of \SI{400}{ms} below 500K spheres, whereas Pulsar remains still below this value at the maximum benchmarked amount of spheres at around 4.4M spheres.
Apart from good scaling behavior in terms of number of spheres, we also observe good scaling behavior in terms of image size.
For 4K image resolution with 4.4M spheres, we still measure execution times of less than \SI{400}{ms} depending on GPU and scene: we measure \SI{387}{ms} for 4.4M spheres on an NVIDIA RTX 2070 GPU) with moderate memory requirements (\SI{3500}{MB}).
\label{sec:render_eq}

\begin{listing}
\begin{minted}
[
frame=lines,
framesep=2mm,
baselinestretch=0.8,
fontsize=\footnotesize,
linenos,
numbersep=6pt
]
{python}
import torch; torch.manual_seed(1)
from pulsar import Renderer
n_spheres = 10
# We create a renderer for a 1024x1024 image.
renderer = Renderer(1024, 1024, n_spheres)
pos = torch.rand(n_spheres, 3) * 10.0
pos[:, 2] += 25.0; pos[:, :2] -= 5.0
col = torch.rand(n_spheres, 3)
rad = torch.rand(n_spheres)
cam = torch.tensor(
  #------t------  ------R------   f    s
  [0.0, 0.0, 0.0, 0.0, 0.0, 0.0, 5.0, 2.0],
  dtype=torch.float32)
image = renderer(pos, col, rad, cam,
  gamma=0.1, max_depth=45.0)
# L1 loss, assuming `target` is an image.
# loss = (image - target).abs().sum()
# loss.backward()
# Use any PyTorch optimizer for optimization.
\end{minted}
\vspace*{-0.5cm}
\caption{Full code example to render a minimal scene with 10 random spheres. The loss computation is shown in ll. 16 and 17 and can be used in any PyTorch optimization loop. $f$ and $s$ denote focal length and sensor width.\vspace*{-0.5cm}}
\label{lst:example}
\end{listing}

\paragraph{Ease of Use}
Lst.~\ref{lst:example} shows a full, executable example for generating and rendering a scene representation in only nine lines of code.
The resulting image will contain ten spheres, randomly placed and colored.
On construction, the renderer creates several workspace and buffer structures, for which the image size and the maximum amount of spheres to render must be known (l. 5).
As you can see in ll. 6-8, the scene representation is encoded in an intuitive way that can be trivially integrated into a PyTorch \texttt{nn.Module} structure or generated through other operations.
Similarly, the camera parameters are encoded in an optimization-friendly format (Pulsar accepts 6D rotation vectors~\cite{zhou2019continuity} natively for better rotation gradients) so that the camera parameter vector can directly be used in a PyTorch optimizer (l. 9).
In l. 13, the render forward function is executed to generate an image tensor from the scene description.
This tensor can be used for further computations, for example neural shading, or directly be used to define a loss (illustrated in ll. 16 and 17).
In either case, the resulting chain of operations is automatically registered with PyTorch's autograd system.

\begin{figure*}
    \includegraphics[width=0.98\textwidth]{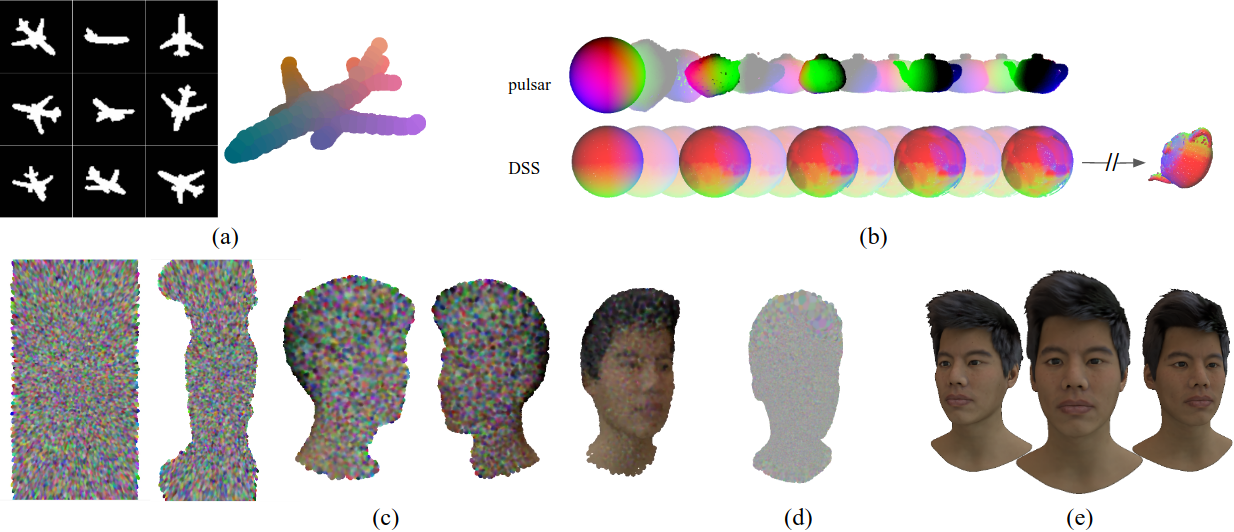}
    \vspace*{-0.3cm}
    \caption{3D reconstruction with Pulsar with up to \num{400}k spheres. \textbf{(a)}~Silhouette-based deformation reconstruction (c.t.~\cite{liu2019soft}); \num{1352} spheres, $64\times 64$. \textbf{(b)}~Reconstruction with lighting cues and comparison with DSS~\cite{yifan2019differentiable}; \num{8003} spheres, $256\times 256$. Pulsar finishes the reconstruction after \SI{31}{s}, whereas DSS finishes after \SI{1168}{s}. \textbf{(c)}~Reconstruction steps of a 3D head model in \SI{73}{s}; \num{400}k spheres, $800\times 1280$. 80 images with random azimuth and elevation are used. \textbf{(d)}~Initialized features for training the neural shading model for (e). \textbf{(e)}~Neural rendering results of a pix2pixHD~\cite{wang2018pix2pixHD} model based on this geometry.}
    \label{fig:recon-low-res}
    \vspace*{-0.4cm}
\end{figure*}

\section{Applications}
We demonstrate several applications in 3D reconstruction and neural rendering to show the versatility of Pulsar.
In particular, we show how Pulsar makes it possible to attack different classical computer vision tasks in a straightforward way with its ability to use a highly detailed representations and high quality gradients.
In all experiments we optimize or \textit{reconstruct} appearance and geometry solely using a photometric $\ell_1$-loss.
We order the experiments by complexity, and in the last presented ones we use several millions of spheres during the optimization.
These experiments would be prohibitively slow to perform with other differentiable renderers.

\subsection{3D Reconstruction}

\paragraph{Silhouette-based 3D Reconstruction}
Pulsar enables 3D reconstruction of objects based on silhouettes.
We demonstrate results on an example scene of SoftRas~\cite{liu2019soft} that provides 120 views of an airplane, see Fig.~\ref{fig:recon-low-res}(a).
To work with our sphere-based representation, we place spheres at all vertices of the mesh SoftRas employs for initialization.
SoftRas uses an image resolution of size $64\times 64$, which pushes the size of each sphere to the lower limits in terms of pixel size.
Instead of the more intricate and computationally complex IOU, Laplacian, and flattening losses that are required in SoftRas, we solely employ a photometric $\ell_1$-loss with respect to the ground truth silhouettes.
SoftRas requires these additional losses to keep the mesh surface consistent.
In contrast, we can move spheres without taking surface topology into account.
For a low number of spheres and small resolution SoftRas is faster, but Pulsar scales much better to real world scenarios (see Tab.~\ref{tab:runtimes}).

\begin{figure*}
    \includegraphics[width=0.98\textwidth]{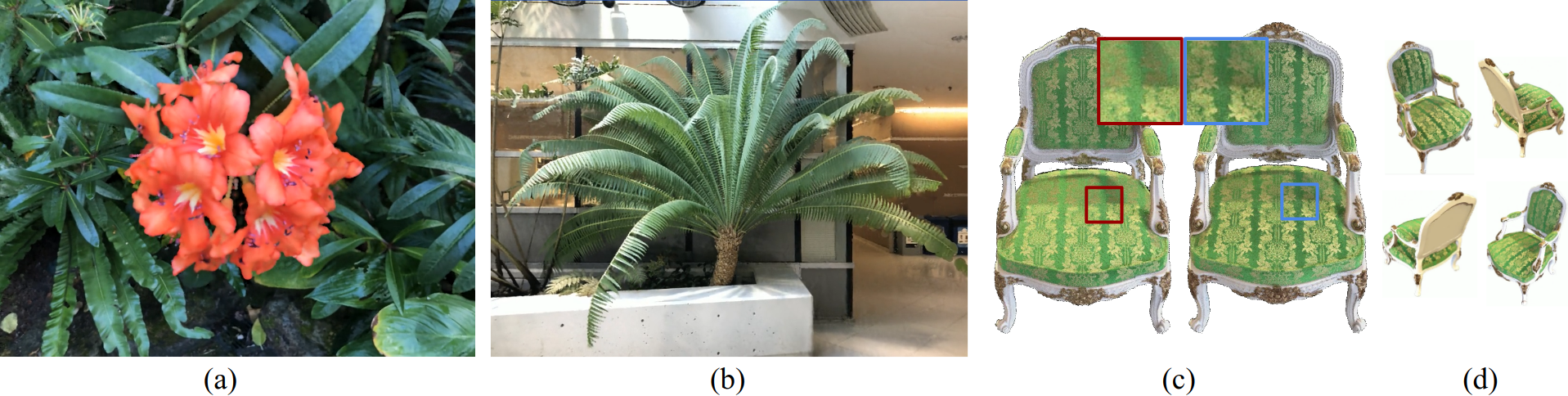}
    \vspace*{-0.3cm}
    \caption{
    High-resolution scene representation view synthesis and reconstruction examples with 1M and more spheres; scenes from the NeRF dataset~\cite{mildenhall2020nerf}. \textbf{(a)}~Test view of the `flower' scene; 2.1M/810K spheres; $1008\times 756$. \textbf{(b)}~Test view of the `fern' scene; 2.6M spheres; $1008\times 756$. \textbf{(c)}~Test view of the `chair' scene with two different virtual viewpoints and shared per-pixel fully-connected shading model;  5.5M/509K spheres, $1600\times 1600$. Note the viewpoint-dependent shading effects on the chair cover. \textbf{(d)}~360 degree views of the `chair' model. X/Y spheres are before/after optimization.}
    \vspace*{-0.5cm}
    \label{fig:recon-high-res}
\end{figure*}

\paragraph{Illumination Optimization}
We reproduce an experiment from DSS~\cite{yifan2019differentiable} that includes an illumination model, see~Fig.~\ref{fig:recon-low-res}(b).
To this end, we implement diffuse shading with parallel light sources as a separate stage.
This highlights the versatility of a dedicated geometry projection step and demonstrates how easy it can be combined with additional refinement models.
Similar to DSS, we use \num{144} cameras, selected at random azimuth and elevation angles, with a constant distance to the object center.
In this experiment, \num{300} optimization steps suffice for Pulsar to reach convergence.
Using Pulsar, we complete the optimization in 31 seconds, whereas DSS requires more complex losses and almost 20 minutes to converge after 477 steps; Pulsar is more than one order of magnitude faster.

\paragraph{Detailed 3D Reconstruction}
Pulsar can go far beyond the number of spheres and image resolutions in the previous examples.
We demonstrate this by reconstructing a head model with realistic hair and eyes at high resolution, see~Fig.~\ref{fig:recon-low-res}(c), from 100 images of resolution $800\times 1280$ pixels.
We initialize a volume with \num{400}k randomly distributed spheres and optimize for a coarse head model in only \SI{73}{s}.
We eliminate spheres if their color converges towards the background color or if they are not visible (spheres that do not receive gradient updates from any viewpoint).
This results in a hull representation of the head with a thickness of several centimeters.
After the optimization and cleanup, a model consisting of approx.~\num{20}k~spheres remains.
Next, we increase the number of spheres three times through subdivision:
We refine each sphere with 12 spheres with radius $\sqrt{2}\cdot r$, where $r$ is the previous radius, and place them in a face-centered cubic packing scheme, re-optimize and prune.
The final model has $~$\num{130}k spheres, see~Fig.~\ref{fig:recon-low-res}(d).
The refinement finishes in \SI{37}{minutes} and temporarily produce models with up to \num{1.6} million spheres.

\paragraph{Neural Shading}
To showcase the potential of the proposed pipeline, we combine the learned head model, see Fig.~\ref{fig:recon-low-res}(d), with a neural shading network for modelling the face in high resolution.
As architecture, we used an off-the-shelf Pix2PixHD~\cite {wang2018pix2pixHD} design and used 15 feature channels.
For the neural network training, we employ a photometric $\ell_1$-loss, a perceptual loss \cite{simonyan2014very}, and an adversarial loss, as is standard for Pix2PixHD.
We experimented with different number of training images: with 80 images in the training set we can already obtain a reconstruction that interpolates well between perspectives, but still with a visible loss in detail.
With more than \num{320} images there's hardly any perceptual difference between training and validation results visible any more, see Fig.~\ref{fig:recon-low-res}(e).
The model produces compelling results that can be rendered in near real time on consumer hardware (we achieve 30+ FPS for the geometry projection step and neural shading takes \SI{37}{ms}).

\subsection{Neural Rendering}
\paragraph{Novel-view Synthesis on Real Data}
We perform novel-view synthesis experiments on examples of the NeRF dataset~\cite{mildenhall2020nerf}.
In the first example, see Fig.~\ref{fig:recon-high-res}(a), we show that pure sphere geometry is sufficient to represent complex, real-world scenes.
We initialize the scene by filling the volume in front of the camera uniformly with 2.1M spheres, with increasing radius according to depth.
We use Pulsar to jointly optimize all sphere properties (i.e., position, radius, color and opacity) based on an $\ell_1$-loss using the Adam optimizer.
We apply a threshold to spheres with low opacity and use an opacity-depth regularizer with the energy $-z_i\cdot O_i$ to encourage spheres to move to the right scene depth. The optimization is completed in 20 minutes on an NVIDIA V100 GPU.
The second example, see Fig.~\ref{fig:recon-high-res}(b), shows a similar reconstruction on the `fern' scene.
We start the optimization with 5.5M random spheres and optimize across three scales to account for the details in the fern leafs.

\vspace{-0.5cm}
\paragraph{View-dependent Shading}
We show separation into geometry projection and neural shading under challenging conditions, see Fig.~\ref{fig:recon-high-res}(c-d).
For this, we consider the synthetic `chair' scene of the NeRF dataset, which has 200 views.
The surface is highly textured and changes its appearance dramatically (satin cover) depending on the viewing angle.
We start the optimization from 5.5M randomly initialized spheres and work in double image resolution to capture all texture details.
We add a simple fully-connected model that is shared across all pixels to optimize for view-dependent appearance and condition on a per-pixel view direction.
Gradients are back-propagated through the shading model to optimize the channel information.
Even such a small model captures the viewpoint dependent effects well.

\section{Limitations}
While we have demonstrated high performance differentiable rendering of complex scenes with millions of spheres, our approach is still subject to a few limitations that can be addressed in future work:
1) It is challenging to compute gradients with respect to the position and radius of spheres that are smaller than a few pixel in screen space.
We address this by explicitly handling this case in the rendering pipeline and by limiting the gradient computation for these spheres to the feature channels and opacity.
In this way, they can be pruned through the opacity optimization and we prevent noisy gradients from leaking into the position- or radius-models.
By finding better ways to handle these cases, we could obtain even better results.
2) While our renderer is highly modular and can be easily integrated with arbitrary PyTorch training loops, it is currently not programmable.
The CUDA kernels are highly optimized and contain the blending function and its symbolic derivative.
This means that changing the function requires explicitly modifying the CUDA kernels, which is time consuming and error prone.
A programmable shader language in combination with auto-differentiation could alleviate this problem, while maintaining the high performance of hand written CUDA code.
\section{Conclusion}
We presented Pulsar, an efficient sphere-based differentiable renderer.
Its architecture builds on recent insights in the fields of differentiable rendering and neural networks and makes deliberate choices to limit complexity in the projection step to obtain high speed and scalability.
Pulsar executes orders of magnitude faster than existing techniques and for the first time enables real-time rendering and optimization of representations with millions of spheres.
We demonstrated its performance and flexibility on a large variety of applications ranging from 3D reconstruction to general neural rendering.
Pulsar is open-source software, modular, and easy-to-use due to its tight integration with PyTorch.
Through its performance and accessibility, we hope that Pulsar will enable researchers to explore new ideas that were out of reach before.

\twocolumn[{%
\renewcommand\twocolumn[1][]{#1}%
\begin{center}
    \Large \textbf{Appendix}
\end{center}%
}]

\appendix
\section{Prelude}

In the appendix, we delve deeper into the implementation details of the Pulsar renderer.
For this purpose, let's briefly revisit the scene description as described in the main paper.
We represent the scene as a set $\mathcal{S}= \{(\mathbf{p}_i, \mathbf{f}_i, r_i, o_i)\}_{i=1}^M$ of $M$ spheres with learned position $\mathbf{p}_i \in \mathbb{R}^3$, neural feature vector $\mathbf{f}_i\in \mathbb{R}^d$, radius $r_i\in \mathbb{R}$, and opacity $o_i\in \mathbb{R}$.
Pulsar implements a mapping $\mathbf{F} = \mathcal{R}(\mathcal{S}, \mathbf{R}, \mathbf{t}, \mathbf{K})$ that maps from the 3D sphere-based scene representation $\mathcal{S}$ to a rendered feature image~$\mathbf{F}$ based on the image formation model defined by the camera rotation $\mathbf{R}$, translation $\mathbf{t}$, and intrinsic parameters $\mathbf{K}$. $\mathcal{R}$ is differentiable with respect to $\mathbf{R}, \mathbf{t}$ and most parts of $\mathbf{K}$, \ie, focal length and sensor size.

\paragraph{The Aggregation Function}
The rendering operation $\mathcal{R}$ has to compute the channel values for each pixel of the feature image $\mathbf{F}$ in a differentiable manner.
To this end, we propose the following blending function (Eq.~1 of the main paper) for a given ray, associating a blending weight $w_i$ with each sphere $i$:
\begin{align}
    w_i = \frac{o_i \cdot d_i \cdot \textcolor{Black}{\exp}(o_i\cdot\frac{\textcolor{Black}{z_i}}{\gamma})}{\exp(\frac{\epsilon}{\gamma})+\textcolor{Black}{\sum_k} o_k\cdot d_k\cdot\textcolor{Black}{\exp}(o_k\cdot\frac{\textcolor{Black}{z_k}}{\gamma})}.
    \label{eq:blending}
\end{align}
We employ normalized device coordinates $z_i \in [0, 1]$ where 0 denotes maximum depth and $d_i$ is the normalized orthogonal distance of the ray to the sphere center.
This distance, since always orthogonal to the ray direction, automatically provides gradients for the two directions that are orthogonal to the ray direction. Strictly speaking, for one ray this direction gradient could be non-existent if the ray hits the sphere in its center; or it could just provide gradients in one of the two remaining directions if it hits the sphere perfectly above or to the side of its center. We provide position gradients only for spheres that have more than three pixels projected radius because we observed that the gradients are numerically not stable otherwise. This means, that \emph{if} position gradients are provided they can move spheres in all directions in space.
We define $d_i = \min(1, \frac{||\vec{d}_i||_2}{R_i})$, where $\vec{d}_i$ is the vector pointing orthogonal from the ray to the sphere center.
Like this, $d_i$ becomes a linear scaling factor in $[0, 1]$.

A notable property of the proposed aggregation function is, that it is commutative w.r.t. the sphere order. This will become important in the following sections.

\section{Data-parallel Implementation}

Modern GPU architectures offer a tremendous amount of processing power through a large number of streaming multiprocessors and threads, as well as enough memory to store scene representations and rendered images in GPU memory. For example, even an NVIDIA RTX 2080 Ti consumer GPU has \num{4352} CUDA cores with \num{64} streaming multiprocessors with access to up to \SI{11}{GB} of memory. The CUDA cores/threads are grouped in \emph{warps} of 32 threads. Multiple warps again can work together in \emph{groups}. Warps have particularly fast local shared memory and operations, however all threads in a warp execute exactly the same command on potentially different data, or a part of them must sleep.
This computing paradigm is called `Single instruction, multiple data' (SIMD).
For example, if half of the threads follow a different execution path due to an `\texttt{if}' statement than the rest; in this case the half not following the branch will sleep while the first half executes the relevant commands.

All these architectural peculiarities must be taken into account to use GPU hardware efficiently. This requires making smart use of parallel code and finding good memory access patterns to not block execution through excessive IO loads. Because both of these aspects tightly connect, non-intuitive solutions often turn out to be the most efficient and experimentation is required to identify them.

We found a way to keep the computation throughput high by elegantly switching between parallelization schemes and by using finely tuned memory structures as `glue' between the computations. In the following sections, we aim to discuss these steps, the underlying memory layout and the parallelization choices made.

\subsection{The forward pass}

For the forward pass, the renderer receives a set of $n$ spheres with position $\mathbf{p}_i$, features $\mathbf{f}_i$, radius $r_i$ and opacity $o_i$ for each sphere $i\in {1, \ldots, n}$. Additionally, the camera configuration $\mathbf{R}$, $\mathbf{t}$ and $\mathbf{K}$ must be provided. Assuming we have a commutative per-pixel blending function, Eq.~\ref{eq:blending}, the first fundamental choice to make is whether to parallelize the rendering process over the pixels or the spheres.

Parallelizing over the spheres can be beneficial through the re-use of information to evaluate the rendering equation for pixels close to each other. However, this approach leads to memory access collisions for the results (writing access to all pixel values must be protected by a mutex), which obliterates runtime performance. The second alternative is to parallelize rendering over the pixels. To make this strategy efficient, it is critical to find a good way to exploit spatial closeness between pixels during the search for relevant spheres. It is important to reduce the amount of candidate spheres (spheres that could influence the color of a pixel) for each pixel as quickly as possible. This can be achieved by mapping spatial closeness in the image to `closeness' on GPU hardware: thread groups can analyze spheres together and share information. Overall, a two step process becomes apparent: 1)~find out which spheres are relevant for a pixel (group), 2)~draw the relevant spheres for each pixel. Both steps must be tightly interconnected so that memory accesses are reduced to a minimum.

By design of our scene parameterization the enclosing rectangle of the projection of each sphere is simple to find. But even in this simple case we would do the intersection part of the calculation repeatedly: every pixel (group) would calculate the enclosing rectangle for each sphere. This is why we separate the enclosing rectangle computation as step (0) into its own GPU kernel. Importantly, through this separation, we can parallelize step (0) over the spheres and use the full device resources.

\subsubsection{Step 0: Enclosing Rectangle Calculation}

This step is parallelized over the spheres. It uses $\mathbf{K}$, $\mathbf{p}_i$ and $r_i$ to determine the relevant region in the image space for each sphere and to encode the intersection and draw information in an efficient way for the following steps. The standard choice for such an encoding is a $k$-d-tree, bounding volume hierarchy (BVH) or a similar acceleration structure. We experimented with (extended) Morton codes~\cite{morton1966computer,vinkler2017extended} and the fast parallel BVH implementations~\cite{karras2012maximizing,karras2013fast} and found their performance inferior\footnote{We used our own implementation that closely follows Karras et al.'s papers but is likely slower than theirs. We evaluated the patented tr-BVH implementation in the NVIDIA OPTIX package (\url{https://developer.nvidia.com/optix}). However, OPTIX does not provide access to the acceleration structure and just allows to query it. This is insufficient for our use case because we need to find an arbitrary number of closest spheres to the camera; we decided not to use OPTIX to avoid the runtime hit of manual sorting.} compared to the following strategy using bounding box projections.

Instead of using acceleration structures, the sphere geometry allows us to find the projection bounds of the sphere on the sensor plane. This is done with only a few computations for the orthogonal but also the pinhole projection model. In the pinhole model, the distortion effects make slightly more complex computations necessary; trigonometric functions can be avoided for higher numerical accuracy through the use of several trigonometric identities.

Additional steps must be taken to robustify the calculated boundaries for numerical inaccuracies. We make the design choice to have every sphere rendered with at least a size of one pixel: in this way, every sphere always receives gradients and no spheres are `lost' between pixel rays.
We store the results of these calculations in two data structures:

\begin{description}
    \item[Intersection information] This is a \texttt{struct} with four \texttt{unsigned short} values and contains the calculated $x$ and $y$ limits for each sphere. This data structure needs 8 bytes of memory. One cache line on the NVIDIA Turing GPUs holds $256=8\cdot 32$~bytes, meaning that all 32 threads in a warp can load one of these data structures with one command. This makes coalesced iteration fast, which helps to process large amounts of intersection data structures in parallel.
    \item[Draw information] This is a \texttt{struct} with all the information needed to draw a sphere once an intersection has been detected. We store the position vector, up to three feature value floats or a float pointer (in case of more than 3 features), as well as the distance to the sphere center and the sphere radius. This requires $8\cdot 4 =$ \SI{32}{bytes} of storage per sphere. The importance of this step is to localize all required information and convert a `struct of arrays' (separate arrays with position, radius, colors) to an `array of structs' (one array of \emph{draw information} structures) with the required information.
\end{description}

After this step, all input variables ($\mathbf{p}_i$, $\mathbf{f}_i$, $r_i$ and $o_i$) are encoded in these two datastructures and only used from these sources.  The computation and storage run in \SI{0.22}{ms} for \num{1000000} spheres.
We additionally store the earliest possible intersection depth for each sphere in a separate array. For spheres that are outside of the sensor area, this value is set to infinity. Then, we use the CUB library\footnote{\url{http://nvlabs.github.io/cub/}} to sort the \emph{intersection information} and \emph{draw information} arrays by earliest possible intersection depth. This step takes another \SI{3.2}{ms} for \num{1000000} spheres. The sorting is important for the following steps: the sphere intersection search may be stopped early once it has been determined that no sphere with a greater distance can still have a notable impact.

\subsubsection{Step 1: Intersection Search}

The aim for this step is to narrow down the number of spheres relevant for pixels at hand as much and as quickly as possible, leveraging as much shared infrastructure as possible. That's why in a first processing step, we divide the entire image into nine parts\footnote{During sorting, we also find the enclosing rectangle for all visible spheres and use this information for tighter bounds of the region to draw.} (an empirically derived value). The size of each of the nine parts is a multiple of thread block launch sizes (we determined this to be $16\cdot 16$ pixels on current GPU architectures). All nine parts are processed sequentially. For each part, we first use the full GPU to iterate over the \emph{intersection information} array to find spheres that are relevant for pixels in the region (we can again parallelize over the spheres). Using the CUB \texttt{select\_flags} routine, we then quickly create arrays with the sorted, selected subset of \emph{intersection information} and \emph{draw information} data structures for all spheres (important: the spheres in this selected subset are still \emph{sorted by earliest possible intersection depth}). From this point on, we parallelize over the pixels and use blocks and warps to use coalesced processing of spheres.

The next level is the block-wise intersection search. We use a block size of $16 \cdot 16 = 256$ threads, so eight warps per block. We observed that larger block sizes for this operation always improved performance, but reached a limit of current hardware at a size of 256 due to the memory requirements. This indicates that the speed of the proposed algorithm will scale favorably with future GPU generations.

We implement the intersection search through coalesced loading of the \emph{intersection information} structures and testing of the limits of the current pixel block. The sphere \emph{draw information} for spheres with intersections are stored in a block-wide shared memory buffer with a fixed size. This size is a multiple of the block size to always be able to accommodate all sphere hits. Write access to this buffer needs to be properly synchronized. If the buffer becomes too full or the spheres are exhausted, Step 2 execution is invoked to clear it. In Step 2, each pixel thread works autonomously and care must be taken to introduce appropriate synchronization boundaries to coordinate block and single thread execution. Additionally, each pixel thread can vote whether it is `done' with processing spheres and future spheres would have not enough impact; if all pixels in a block vote `done', execution is terminated. The vote is implemented through a thread-warp-block stage-wise reduction operation.

\subsubsection{Step 2: the \emph{Draw} Operation}

The draw operation is executed for each pixel separately and for each sphere \emph{draw information} that has been loaded into the shared memory buffer. Because every pixel is processed by its own thread, write conflicts for the channel information are avoided and each pixel thread can work through the list of loaded \emph{draw information} at full speed. The intersection depths for each sphere are tracked: we use a small (in terms of number of spheres to track; this number is fixed at compile time) optimized priority queue to track the IDs and intersection depths of the closest five spheres for the backward pass. Additionally, updating the denominator of the rendering equation allows us to continuously have a tracker for the minimum required depth that a sphere must have for an an $n$ percent contribution to the color channels. If set (default value:~1\%), this allows for early termination of the raycasting process for each pixel.

\subsubsection{Preparing for the Backward Pass}

If a backward pass is intended (this can be optionally deactivated), some information of the forward pass is written into a buffer. This buffer contains for each pixel the normalization factor as well as the intersection depths and IDs of the closest five spheres hit.

We experimented with various ways to speed up the backward calculation and storing this information from the forward operation is vastly superior to all others. It allows to skip the intersection search at the price of having to write and load the backward information buffer. Since writing and loading can be performed for each thread without synchronization, it turned out to be the most efficient way.

\subsection{The Backward Pass}

Even with the intersection information available, there remain multiple options on how to implement the backward pass. It is possible to parallelize over the spheres (this requires for each thread to iterate over all pixels a sphere impacts, but it avoids synchronization to accumulate gradient information) or over the pixels (this way each thread only processes the spheres that have been detected at the pixel position, but requires synchronization for gradient accumulation for each sphere). We found that parallelizing over the pixels is superior, especially since this implementation is robust to large spheres in pixel space.

Again, minimizing memory access is critical to reach high execution speeds. To achieve this, we reserve the memory to store all sphere specific gradients. Additionally, we also allocate a buffer for the camera gradient information \emph{per sphere}. We found that accumulating the sphere gradients through synchronized access from each pixel thread is viable, but synchronizing the accumulation of the camera gradients, for which every sphere for every pixel has a contribution, causes too much memory pressure. Instead, we accumulate the camera gradients sphere-wise and run a device-wide reduction as a post-processing step. This reduces the runtime cost to only \SI{0.6}{ms} for \SI{1000000} spheres.

Overall, this implementation proved robust and fast in a variety of settings. Apart from being nearly independent of sphere sizes, it scales well with image resolution and the number of spheres. We found additional normalization helpful to make the gradients better suited for optimization:
\vspace*{-0.6cm}
\begin{itemize}
    \item sphere gradients are averaged over the number of pixels from which they are computed. This avoids parameters of small spheres converging slower than those of large spheres. In principle, large spheres have a larger impact on the image, hence receive larger gradients. However, from an optimization point of view, we found the gradients normalized by the number of pixels are much better suited for stable loss reduction with gradient descent techniques.\vspace*{-0.2cm}
    \item camera gradients need to take the sphere size into account to lead to a stable optimization. We use the area that each sphere covers in the image as a normalization factor (together with the constant \num{1e-3}, which we found a suitable normalization factor in practice). The area normalization makes this calculation very similar to Monte Carlo integration.
\end{itemize}

\noindent The gradient computation for each of the gradients is only performed if the gradients are required by the PyTorch autodiff framework. Overall, using these strategies we achieve very good scaling behavior as demonstrated in Fig.~\ref{fig:scaling}.

{
\small
\bibliographystyle{ieee_fullname}
\bibliography{egbib}
}

\end{document}